\newcommand{\citepeg}{\citep[e.g., ][]}

\providecommand{\adsurl}[1]{\href{#1}{ADS}}

\documentclass[iop,twocolumn]{aastex63}

\pdfoutput=1 
\usepackage{amsmath,amstext}
\usepackage[T1]{fontenc}
\usepackage{apjfonts} 
\usepackage{graphics,graphicx,epsf,natbib}
\usepackage{subfigure}

\shorttitle{Clusters Masquerading as X-ray Quasars}
\shortauthors{Donahue et al.}

\begin{document}

\title{Clusters of Galaxies Masquerading as X-ray Quasars}

\author{Megan Donahue, Kelsey Funkhouser, Dana Koeppe, Rachel L. S. Frisbie, G. Mark Voit}
\affiliation{Michigan State University, Department of Physics \& Astronomy, 567 Wilson Road, East Lansing, MI 48824 USA}

\begin{abstract}

Inspired by the discovery of the Phoenix cluster by the South Pole Telescope team, we initiated a search for other massive clusters of galaxies missing from the standard X-ray catalogs. We began by identifying 25 cluster candidates not included in the Meta-Catalog of X-ray Clusters of galaxies cluster compilation through cross-identification of the central galaxies of optically identified clusters in the Sloan Digital Sky Survey GMBCG catalog (Hao et al. 2010) with bright X-ray sources in the ROSAT Bright Source Catalog. Those candidates were mostly unidentified or previously classified as X-ray active galactic nucleus (AGN). We analyzed brief {\em Chandra} X-ray Observatory observations of 14 of these X-ray sources and found that eight are X-ray luminous clusters of galaxies, only one showing evidence for a central X-ray point source. The remaining six candidates turned out to be point-source dominated, with faint detections or upper limits on any extended emission. We were not able to rule out the presence of extended X-ray emission from any of the point sources. The levels of extended emission around the six point sources are consistent with expectations based on optical richness, but could also be contaminated by scattered X-ray light from the central point source or extended nonthermal emission from possible radio lobes. We characterize the extended components of each of the well-detected cluster sources, finding that six of the eight X-ray clusters are consistent with being compact cool-core clusters. One of the newly identified low-luminosity X-ray clusters may have had an X-ray-luminous AGN 20 yr prior to the recent {\em Chandra} observations, based on the 4-$\sigma$  difference between its {\em Chandra} and ROSAT fluxes.

\end{abstract}


\section{Introduction and Scientific Motivation}
The discovery of the Phoenix Cluster \citep{Williamson_2011,2012Natur.488..349M}, the most luminous X-ray cluster known, rocked the worldview of X-ray astronomers. 
Stunningly, this remarkable X-ray cluster was identified not with an X-ray telescope but rather in the microwave band with the South Pole Telescope (SPT; \citet{2011PASP..123..568C}). X-ray follow-up by Michael McDonald and his team showed that it is an extraordinarily massive ($\sim 2 \times 10^{15}$ M$_\odot$) cluster at $z = 0.597$.  Its spectacular Brightest Cluster Galaxy (BCG) hosts a tremendous starburst ($\sim  740 ~M_\odot$ yr$^{-1}$) and a central X-ray point source surrounded by hot, dense, and luminous X-ray plasma \citep{2012Natur.488..349M,McDonald_2013,2019ApJ...885...63M}.  Massive clusters of galaxies are distinctive and luminous X-ray sources because most of their baryons are in the form of hot, diffuse, volume-filling intergalactic plasma,  with a gas temperature characteristic of the gravitational potential well ($T \propto M/R \sim 1-15$ keV).
When are enormously luminous X-ray clusters discovered and correctly identified with radio telescopes?  The answer is: when they were originally classified as bright X-ray point sources associated with a previously known active galactic nucleus (AGN).  The Phoenix Cluster was identified as a bright X-ray source in the ROSAT Bright Source Catalog \citep{Voges_1999} but was classified as an X-ray point source, cross-identified with an optical AGN, and left to languish in AGN anonymity, waiting for its  ``cluster-ness''  to be revealed over a decade later by the SPT survey. 

Clusters like the Phoenix Cluster belong to a relatively common class known as ``cool-core clusters."  Such clusters have prominent, high-X-ray surface brightness central regions (cores) with characteristic radii $\sim 10-30$ kpc  \citep[e.g., ][]{Fabian1994}. These clusters may occasionally have a central X-ray AGN, but such luminous point sources are rare in cool-core clusters. Only about 20\% of {\em Chandra} archival clusters show any evidence at all for a central X-ray AGN \citep{Yang_2018}.  The cool-core phenomenon is more common, occurring in one-third to one-half of all X-ray clusters  \citep{1992ApJ...385...49D, Edge1992,1998MNRAS.298..416P, BauerFabian2005}. 

X-ray selected cluster samples are biased toward cool-core clusters because their bright, high-contrast central emission makes them easier to detect in X-ray imaging surveys \citep{Pesce1990}. A cluster of galaxies more distant than about $z\sim0.1$ has a bright core that may be no larger than $r\sim5\arcsec-15\arcsec$. This angular scale is just below the angular range in which the workhorse detector on board ROSAT \citep{ROSAT1982}, the Position Sensitive Proportional Counter (PSPC; \citealt{2003NIMPA.515...65P}), could resolve it, particularly in the shallow exposures characteristic of the ROSAT All-Sky Survey (RASS;\citealt{1993Sci...260.1769T}).  Furthermore, an X-ray source may also coincide with a luminous optical AGN, which can confound ground-based optical spectroscopic classification if the AGN is not supplying most of the X-ray flux. Such sources, including both very compact, luminous cool-core clusters and X-ray clusters with a luminous central X-ray AGN, may have been systematically misidentified in general X-ray classification efforts. X-ray AGNs are much more common than X-ray clusters, making it unsurprising that some groups and clusters of galaxies may have been misidentified as quasars and AGNs, given the practical limitations of follow-up efforts.

Because clusters of galaxies are so rare, and massive clusters the rarest of all, to find the most massive X-ray clusters in the universe, astronomers must conduct wide-area searches. 
The ROSAT All-Sky Survey \citep{1993Sci...260.1769T,Voges1993} is currently the most recent all-sky X-ray survey and produced valuable catalogs of X-ray clusters 
\citep{NORAS_II,Ebeling1998,MACS2007,MACS2010,REFLEX2004}. There were deeper surveys, some conducted using pointed observations by ROSAT itself
\citep{1997ApJ...477...79S,Mullis160SD,1998ApJ...502..558V,Burenin400SD,1998ApJ...492L..21R,2002ApJ...569..689D}. 
 
The object of the study described in this paper was to unmask X-ray-luminous, massive galaxy clusters that may have been hiding among the bright unresolved sources in the ROSAT All-Sky Survey and to establish the characteristics of this previously hidden population.  We were interested in understanding any selection bias disfavoring these clusters because bias can affect the interpretation of any cluster sample drawn from the ROSAT All-Sky Survey.  X-ray clusters can be used as cosmological probes of the mean density of matter and the amplitude of its initial perturbation spectrum \citep[e.g.,][]{2009ApJ...692.1060V, 2010MNRAS.406.1759M,2008MNRAS.387.1179M}.  Measuring the abundance and evolution of the most massive clusters is extremely important because the properties of massive clusters are exquisitely sensitive to those cosmological parameters \citepeg{Voit_2005,2011ARA&A..49..409A}. If X-ray surveys based on ROSAT (or eROSITA) are missing massive clusters because they look like bright point sources, that incompleteness could affect cosmological conclusions. While none of the X-ray sources we targeted were Phoenix-like in luminosity, all were nonetheless quite luminous, with bolometric $L_x \sim 10^{44} - 2 \times 10^{45}$ erg s$^{-1}$. 

Another motivation for this work arose from questions about galaxy evolution. The extraordinary atmosphere of a cool-core cluster appears to be stabilized by physical processes associated with AGNs (i.e. ``AGN feedback''). In simulations, AGN feedback prevents catastrophic cooling (the ``over-cooling'' problem) by regulating the gas entropy in the cluster core;  higher entropy gas is less susceptible to catastrophic cooling \citep[e.g., ][]{2007MNRAS.380..877S}.  One critical test of AGN feedback implementations in cosmological simulations is whether the simulation can reproduce the observed incidence of cool-core clusters. The cool-core fraction may change with redshift, but selection bias plays an important role in correctly interpreting the observations. A proper observationally based census of how many clusters host a prominent single BCG or a central AGN or are in a specific dynamical state provides a baseline for assessing our models and simulations. Specifically, we now know that the cool-core phenomenon in clusters provides crucial insights into how AGN jets and winds regulate the gas reservoirs that feed both AGN growth and star formation in AGN hosts.  We know that clusters with cool cores tend to host radio-loud AGNs that inflate cavities in the intracluster gas and presumably heat it as well \citep{2007ARA&A..45..117M,2012NJPh...14e5023M}.  
ROSAT studies suggested  \citep{1995AJ....110..513H,1999MNRAS.309..969H}  and XMM and {\em Chandra} studies confirmed \citep{2003MNRAS.339.1163C,2007MNRAS.381.1109B} that show that powerful radio sources likely sit inside X-ray-emitting clusters of galaxies. A deep  {\em Chandra} observation by \citet{2010ApJ...722..102S} of a $z=1$ compact steep-spectrum radio source (3C~186) show it is surrounded by a cool-core X-ray atmosphere. Studies of gas halos around radio sources provide important clues for understanding AGN feedback. It is important to measure how that population of cool-core clusters evolves with time, but in order to do that, we need to know what fraction of the strongest cool-core clusters, the ones with the most highly peaked X-ray surface brightness profiles, may be missing from ROSAT-based cluster catalogs because they are masquerading as point sources.

Finally, while rarely found in clusters of galaxies, X-ray-luminous AGNs in BCGs themselves may yield important clues about the early growth phases. NGC~1275 has a Seyfert 1.5 nucleus \citep{2001ApJS..136...61S}. Some other cluster cores are known to have both a strong X-ray AGN and abundant star formation, such as IRAS~09104 at $\sim0.5$ \citep{2012MNRAS.424.2971O}, PKS1229-021 \citep{2012MNRAS.422..590R}, H1821+64 \citep{2010MNRAS.402.1561R}, 3C~186 \citep{2005ApJ...632..110S,2010ApJ...722..102S} and of course the Phoenix Cluster itself \citep{2012Natur.488..349M}.   In order to assess the duty cycle and the role of X-ray-bright (radiatively luminous) AGNs in cluster cores, we would like to know the fraction of BCGs that have bright X-ray AGNs. If very bright AGN outshine the fainter, extended component of the intracluster gas, we may be missing some of the most extreme but important examples of luminous X-ray AGNs in massive halos.

\section{Target Identification and Observational Strategy}
Our search for suitable targets began when we realized that the Phoenix Cluster had been lying in plain view for years in the RASS Bright Source Catalog \citep{Voges_1999}. We suspected there may be more diamonds in that pile.  The diamond mining started with the locations of 55,000+ clusters from the Sloan Digital Sky Survey (SDSS) GMBCG catalog \citep{Hao_2010}, which we cross-correlated with the locations of  18,806 sources in the ROSAT All Sky Survey Bright Source Catalog \citep{Voges_1999}.  We found 50 matches in which a bright RASS X-ray source was located within $10\arcsec$ of a Sloan Brightest Cluster Galaxy from \citet{Hao_2010}.  Some turned out to be famous cool-core clusters like Zwicky 3146 and Abell 1835.  So, in order to determine which of these 50 matches were previously cataloged X-ray clusters, we consulted the Meta-Catalog of X-ray Clusters of galaxies \citep[MCXC;][]{MCXC}, and we also utilized the NASA Extragalactic Database\footnote{The NASA/IPAC Extragalactic Database (NED)
is operated by the Jet Propulsion Laboratory, California Institute of Technology,
under contract with the National Aeronautics and Space Administration.} (NED) to find other positional cross-identifications. 

This procedure narrowed the sample to 25 candidates that were identified as BCGs of optically-selected GMBCG clusters and coincided with bright RASS X-ray sources but did not appear in any existing X-ray cluster catalog. To create this list, we also excluded sources for which we had found literature identifying them as known X-ray clusters, such as 4C 55.16 \citep{1999MNRAS.306..467I}.  We knew in advance that a few of the 25 sources would almost surely be X-ray clusters, because they were associated with BCGs with spectacular emission-line spectra, suggesting that they were very likely at the centers of cool-core clusters. All but two BCGs had spectroscopic redshifts from either the original GMBCG catalog \citep{Hao_2010} or SDSS Data Release 9 
\citep[DR9;][]{2012ApJS..203...21A}. The two BCGs lacking SDSS spectra had photometric redshifts consistent with the spectroscopic redshifts of their closest red companions.  We then proposed and obtained {\em Chandra} observations over the course of two observing cycles [Cycles 15 and 16, Proposal IDs 15800614 and 16800415) for a subsample of 13 of these 25 X-ray cluster candidates. We included one {\em Chandra} archival observation in our study, for a total of 14 {\em Chandra} targets listed in Table~\ref{Table1}.

We classified our targets based on the SDSS spectra of the BCGs, prior to the {\em Chandra} observations (and prior to the analysis of any {\em Chandra} archival observation). Those classifications are listed in Table 1 as AGN, CC, or \replaced{Q}{OQ}.  ``\replaced{Q}{OQ}'' identifies spectra typical of optically-quiescent galaxies with old stellar populations. ``CC'' flags BCG spectra typical of emission-line BCGs in cool-core clusters, with prominent hydrogen recombination lines and bright, low-ionization forbidden lines ([\ion{O}{2}], [\ion{N}{2}], [\ion{O}{1}], [\ion{S}{2}]). Sometimes, this type of spectrum is classified as a ``LINER,'' but the emission is often quite extended. CC is not an official SDSS classification, but this BCG spectral type is very familiar to astronomers who study BCGs in X-ray clusters of galaxies, such as NGC~1275, Abell~1835, or the Phoenix Cluster. Finally, spectra with a significant nonstellar component (power-law continua, broad emission lines) were classified as ``AGN."  Table 1 reports the original ROSAT source classification based on objective prism spectroscopy from the RASS BCS \citep{2003A&A...406..535Z}, which shows that only two of these sources were identified even tentatively as clusters (ID code 31). Half of the sources were unidentified and the rest were tentatively associated with AGNs (ID code 11 or 10) or galaxies (ID 21).
Our approach is generally similar to the one employed by the Clusters Hidden in Plain Sight (CHiPS) team \citep{Somboonpanyakul_2018}, with the main differences in approach that we base our search on an existing SDSS cluster catalog and include all X-ray sources in the ROSAT Bright Source Catalog, regardless of their flux and associated redshift. 

We report our specific observational strategy and data reduction procedure in \S~3. We fit and derive properties of the cluster X-ray sources in \S~4 and give our estimates of and limits on the hot gas luminosities around the dominant AGN sources in \S~5. We did not attempt to characterize the point sources themselves because of the considerable pile-up uncertainties. In deriving angular size distances and luminosity distances, we assume a geometrically flat cosmology of $H_0=70$ km s$^{-1}$ Mpc$^{-1}$ and $\Omega_M=0.3$.

\section{{\em Chandra} Observations and Data Preparation}

Five cluster candidates were observed between 2014 February and August with the Advanced CCD Imaging Spectrometer (ACIS-I) in FAINT mode. An additional eight targets were observed between 2015 March  and 2016 January in VFAINT mode. Pointings of most of the targets were offset by less than an arcminute from the nominal ACIS-I pointing, in order to obtain the maximum point-source function (PSF) quality and to minimize the effect of chip gaps on possibly extended sources. All reductions and analyses were done with the {\em Chandra} Interactive Analysis of Observations (CIAO;  \citealt{2006SPIE.6270E..1VF}). Over the course of the investigation, we used CIAO versions 4.8-4.10 and the corresponding calibration files (CALDB 4.7.0-4.7.8). The results of our analysis were insensitive to the version of CIAO or the CALDB.  Our main observational goals were to characterize the two-dimensional morphologies of these sources, and to measure the X-ray luminosities and temperatures of any extended sources. The sources that turned out to be AGN point sources were affected by pile-up. We estimated the maximum luminosities for the faint extended components around dominant point sources, but we did not attempt to measure their spectral properties.  We did not attempt to characterize the point sources at all, except to consistently model their large-angle scattering features.  The data were reprocessed using the {\em Chandra}  repro script from CIAO. The data were screened for flares, or high background episodes. The CIAO deflare script uses sigma clipping to identify times of high background, and we created a new Good Time Intervals (GTI) list that was used to filter the high background episodes from the event data. The flare-free observation times averaged 98\% of the total requested times (less than a few minutes each observation), so flare screening had negligible impact on the analysis and the data quality.

{\catcode`&=11
\gdef\2003AandA...406..535Z{\citet{2003A&A...406..535Z}}}

\begin{deluxetable*}{lcccllrlrcll}
\tablecaption{X-Ray Targets and Classifications \label{Table1}}
\tablehead{
\colhead{Name }              & \colhead{Spectral}  & \colhead{{\em Chandra}}  & \colhead{ROSAT} & \colhead{$z$ (SDSS)} & \colhead{ROSAT}    & \colhead{$S_{opt}$}  & \colhead{OBSID} & \colhead{Exp Time} & \colhead{$N_H$} & \colhead{$F_{1.4~GHz}$} & \colhead{$L_{1.4\rm{GHz}}$}  \\ 
\colhead{   }     & \colhead{Class }    & \colhead{Class }  & \colhead{Class}            & \colhead{}         & \colhead{(cps)} &  \colhead{}    & \colhead{} &   \colhead{(sec)} & \colhead{($10^{20}~\rm{cm}^{-2}$)} & \colhead{ (mJy)} & \colhead{W~Hz$^{-1}$} } 
\startdata
RXJ~082814.5+415359 & OQ                       & AGN        & - & 0.226    & 0.089     & 14.3                & 17180 & 16433    & 3.68        & 78.7             & 8.44E+24                       \\
RXJ~085451.0+621843 & AGN                     & AGN        & 11    & 0.267    & 0.064     & 12.5                & 16138 & 17487    & 4.66        & 279              & 4.13E+25                       \\
RXJ~090533.4+184011 & CC                      & Cluster      & 31  & 0.1218   & 0.125     & 11.5                & 16139 & 9849     & 3.71        & 13.5             & 4.31E+23                       \\
RXJ~090953.9+310558 & AGN                     & AGN        & 10  & 0.272    & 0.345     & 10.1                & 17178 & 4818     & 2.22        & 62.4             & 9.58E+24                       \\
RXJ~091651.8+523829 & AGN                     & AGN        & 11  & 0.19     & 0.428     & 6.7                 & 17183 & 4062     & 1.56        & 72.6             & 5.55E+24                       \\
RXJ~092710.8+532741 & CC                      & Cluster      &  -  & 0.201    & 0.12      & 13.9                & 17177 & 13900    & 1.88        & 3.6              & 3.07E+23                       \\
RXJ~103035.2+513229 & CC                      & Cluster      &  -  & 0.5183   & 0.062     & 8.5                 & 16137 & 19947    & 1.25        & 175              & 9.03E+25                       \\
RXJ~111908.5+090017 & CC                      & Cluster       & - & 0.3315   & 0.055     & 16.7                & 16140 & 20707    & 3.02        & 19.1             & 4.28E+24                       \\
RXJ~113121.4+333447 & CC                      & Cluster       & - & 0.221    & 0.113     & 13.8                & 17176 & 13644    & 2.08        & 7.6              & 7.80E+23                       \\
RXJ~121510.9+073205 & OQ                       & AGN          & 21 & 0.136    & 0.365     & 5.8                 & 17181 & 6036     & 1.49        & 81.7             & 3.24E+24                       \\
RXJ~135022.2+094007 & CC                      & Cluster+AGN & 31  & 0.1325   & 0.36      & 5.8                 & 14021 & 19780    & 2.23        & 300              & 1.13E+25                       \\
RXJ~144248.5+120042 & AGN                     & AGN          & 21 & 0.163    & 0.873     & 20.4                & 17179 & 2091     & 1.6         & 71.3             & 4.04E+24                       \\
RXJ~145611.1+302108 & OQ                       & Cluster       & - & 0.4156   & 0.063     & 12                  & 16141 & 19200    & 1.65        & $<1$     & $<3.4E+23$                       \\
RXJ~170559.9+365732 & OQ                       & Cluster       & - & 0.278    & 0.065     & 27                  & 17182 & 24747    & 2.52        & $<0.5$   & $<8.0E+22$                      
\enddata
\tablecomments{Spectral classes are AGN (broad lines, strong blue continua), CC for BCGs with low-ionization nebular emission lines common in cool-core cluster BCGs, and \replaced{Q}{OQ} for BCGs with optical spectra characteristic of red and dead stellar populations. The ROSAT Class is from the optical identifications based on objective prism spectroscopy from \2003AandA...406..535Z. Blanks correspond to non-identifications (flags 803, 8, and 0 in the original catalog). The first digit of the other classifications are AGN=1, Galaxy=2, and Cluster=3, with the second digit indicating confidence of the identification, where 0=highly probable, and 1=probable, but possible issues.The SDSS redshift is from the most SDSS Data Release 14. The ROSAT count rate in counts per second (cps) is from the ROSAT Bright Source Catalog \citep{Voges_1999}. $S_{opt}$ is the optical richness parameter from \citet{Hao_2010}. Each observation is associated with a {\em Chandra} ObsID and net exposure time. X-ray spectra were modeled assuming a Galactic column density $N_H$ reported here in units of $10^{20}$ cm$^{-2}$. Rest-frame 1.4 GHz radio fluxes are from FIRST, and corresponding luminosities were calculated based on $H_0=70$ km s$^{-1}$ Mpc$^{-1}$, $\Omega_M=0.3$, flat cosmology assumption, and spectral shape $L_\nu \propto \nu^{-0.7}$. }
\end{deluxetable*}

A simple inspection of raw {\em Chandra} ($0.5-7.0$ keV) images allowed easy identification of sources dominated by an extended emission (cluster of galaxies) or by a point source (AGN; Table~\ref{Table1}).  For every observation, a broadband ($0.5-7.0$ keV) image and a softer ($0.5-2.5$ keV) image were created using the CIAO  fluximage script. We generated a normalized X-ray exposure map for each image, which also allowed us to create a mask of the gaps between the detectors and bad pixels and to correct for small differences in the the spatial nonuniformity in the sensitivity of the final exposure. We used the mask and the normalized exposure map in our modeling procedure to forward-model the observed counts. Bright point sources were identified with CIAO  wavdetect on a broadband flux image of the field. The sources detected by the task  were manually verified,  then  filtered from the reprocessed event files used for spectroscopy. We also masked these regions from the morphological analyses.  We did not exclude central sources. Such sources, if present, were modeled in the morphological analyses. 
For our morphological analyses, we report results from the broad-band data (0.5-7.0 keV) in order to maximize the signal to noise. Because temperature gradients could affect the gradients in a broadband surface brightness image, we repeated the morphological analyses with the softer image to check whether the choice of bandpass mattered. The best-fit parameters and errors emerging from the modeling of the soft X-ray images had very similar, statistically compatible results, so we report only results from the analyses of images created from the broadband data.  Uncertainties reported for any best-fit parameters are $1-\sigma$ estimates. All upper limits are $3-\sigma$.

\section{Extended Sources}

This section discusses the eight targets readily identified as extended X-ray sources with little to no evidence for central point sources. All eight sources are significantly more extended ($r\sim 1'$) than a {\em Chandra} point source ($r \sim 1"$).

\subsection{Spatial Analysis}
Because we are characterizing sources that were previously missed in other X-ray searches, one question was whether these sources were more compact than typical X-ray selected clusters (and thus missed as an extended source). None of these sources apart from the archived observation of RXJ~1350+09 exhibits a visible central point source, so we also conducted an analysis of each image to place an upper limit on the contribution to a central point source. The net number of broad-band counts from the clusters fit to each model was about 3000 in each case, except for RXJ~0905, for which had only 1000 counts because the source apparently varied. (See \S~ \ref{section:0905} for more information.) 

\begin{figure*}
\includegraphics[width=\linewidth, angle=0]{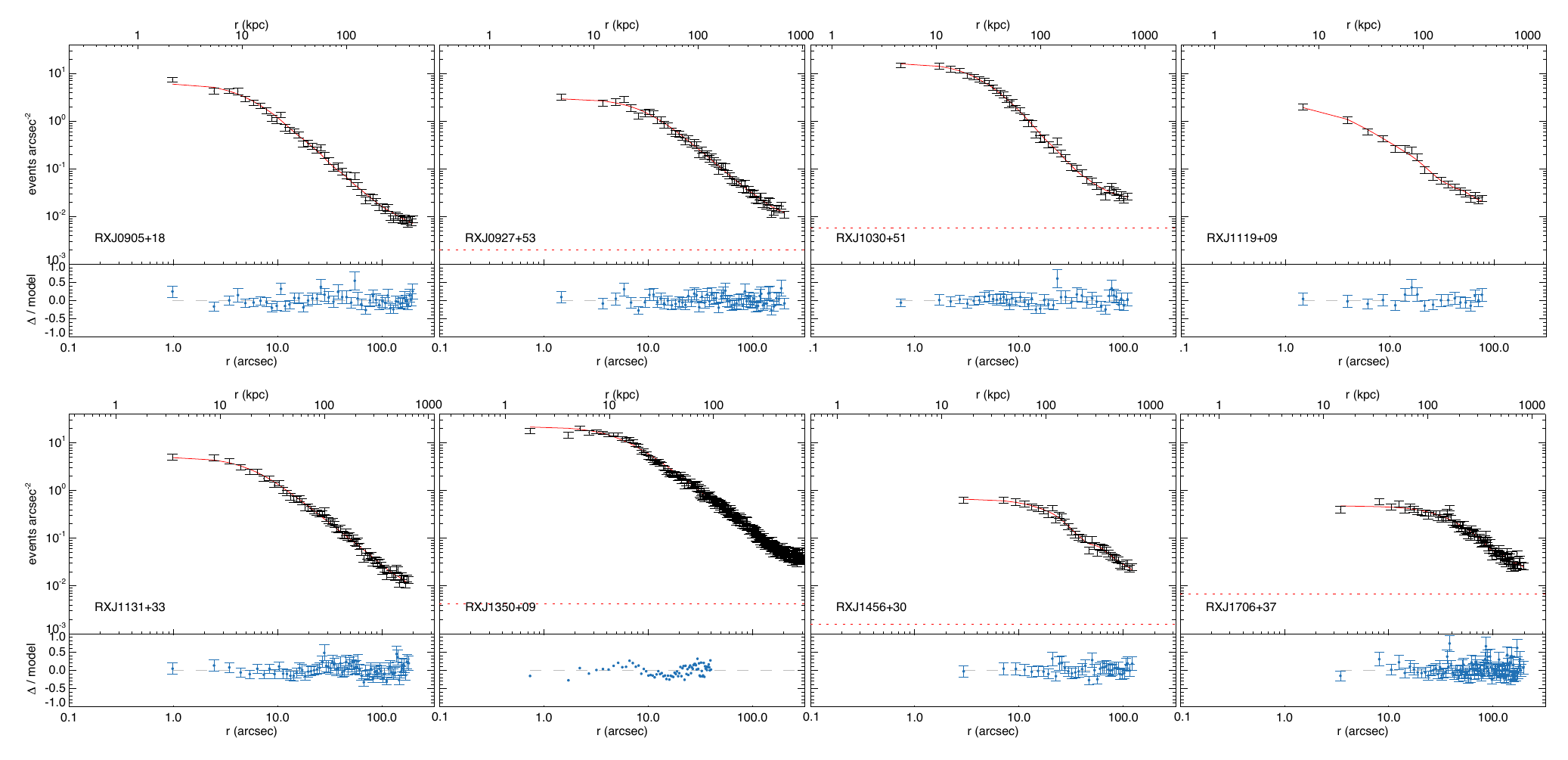}
\caption{
 Radial profiles based on annular bins from the 2D single-pixel binned x-ray images and best-fit 2D models for the X-ray sources dominated by extended emission. 
These graphs show the data (black points with error bars) binned to 50 events per radial bin. These representations are not the basis for the fits to the models; the fits
were conducted on 2D images.  The red line represents the best-fit 2D $\beta$-model binned in annular rings in an identical fashion as the data. The dotted line, where visible, shows the best-fit background level in each case.
\label{figure:1}
  }
\end{figure*}

In order to make a quantitative estimate of the \replaced{extent}{size of the core} of each source, we fit a standard ``beta'' model \citep{Sarazin_1988} and a uniform X-ray background to two-dimensional X-ray images. The model was multiplied by the exposure map described in the previous section, and the resulting model prediction was then compared to the raw counts data. We used the fitting tool provided in the CIAO package called SHERPA \citep{Sherpa_2001} and its standard 2D beta-model. We allowed all parameters in the model to vary to find a best-fit solution. That is, we fit the central $(x,y)$ position, core radius $r_c$, $\alpha$ where $\alpha=3\beta-1/2$ and $\beta$ is the eponymous $\beta$ in the $\beta$-model \citep{Sarazin_1988}, position angle $\theta$, and ellipticity $\epsilon$ where 
\begin{equation}
SB_X(r) = A \left[ 1+ \left(\frac{r}{r_0}\right)^2 \right]^{-\alpha}.
\end{equation}
In this model $r$, $\theta$, and $\epsilon$ are defined so that
$$r(x,y)= \left[ x_1^2(1-\epsilon)^2+y_1^2\right]^{1/2} (1-\epsilon)^{-1}$$
where $x_1=(x-x_0)\cos(\theta)$ and $y_1=(y-y_0)\sin(\theta)$. ($\epsilon=0$ corresponds to perfectly circular isophotes.)  We use the Poisson log-likelihood function (cstat) as the fit statistic. In SHERPA's implementation of this statistic, an approximate goodness-of-fit estimate can be generated in the sense that the observed statistic divided by the number of degrees of freedom is approximately unity for good fits. The fits fairly represent the X-ray surface brightness profiles, within the limits of the detected flux. We use the contributed CIAO routines prof\_data  and prof\_model to generate identically binned radial representations of the 2D data and its best fit. These profiles are displayed in Figure~\ref{figure:1}. The radial profiles should not be construed as the data that were fit; the fitting process was conducted on the 2D image data as described above.

To estimate an upper limit for a putative central point source, we modified the model by adding a simple Gaussian component with a Full Width Half Maximum (FWHM) restricted to 1.0-2.0 pixels. This quantity is consistent with the ideal {\em Chandra} ACIS-I point source on-axis, with a FWHM and a blur of about $0\farcs5$. These properties are consistent with those of faint point sources observed in the same field.  
We required the putative point source to be centered at the same location as the extended source. The upper limits, for sources which yielded amplitudes consistent with zero, are based on the $3-\sigma$ uncertainties on the amplitudes of these Gaussians.

\begin{deluxetable*}{llllllllllll}
\tablecaption{Extended X-ray Source Morphology \label{table:betamodel}}
\tablehead{
\colhead{Name} & \colhead{RA} &  \colhead{Dec} &   \colhead{Pos. unc} &  \colhead{$R_c$} &  \colhead{$R_c$ Err} & \colhead{$\alpha$} & \colhead{$\alpha$ Err} & \colhead{$\epsilon$} & \colhead{$\epsilon$ Err} & \colhead{PA} & \colhead{PA Err}  \\
      &    &          & \colhead{($\arcsec$)} &  \colhead{(kpc)} &  \colhead{(kpc)} & & & & & \colhead{deg E of N} & \colhead{deg}}
 \startdata
RXJ0905 & 09 05 33.5 & +18 40 02.6 & 0.25 & 8.9 & 0.8 & 0.93 & 0.04 & 0.07 & 0.04 & -33 & 17 \\
RXJ0927 & 09 27 10.6 & +53 27 31.2 & 0.48 & 31.8\tablenotemark{*}  & 2.9 & 1.04 & 0.04 & 0.33 & 0.02 & 37 & 2 \\
RXJ1030 &10 30 35.2 & +51 32 32.2 & 0.29 & 27.4 & 1.8 & 1.28 & 0.04 & 0.16 & 0.03 & -21 & 6 \\
RXJ1119 & 11 19 09.0 & +09 00 22.5 & 0.44 & 11.4 & 3.5 & 0.7 & 0.09 & - & - & - & - \\
RXJ1131 & 11 31 20.9 & +33 34 46.5 & 0.29 & 19.6 & 1.9 & 0.92 & 0.03 & 0.08 & 0.03 & -42 & 10 \\
RXJ1350 & 13 50 22.1 & +09 40 11.1 & 0.58 & 12.5 & 0.6 & 0.88 & 0.01 & 0.16 & 0.01 & 9 & 2 \\
RXJ1456 & 14 56 11.2 & +30 21 02    & 0.89 & 108.6\tablenotemark{*}  & 21.6 & 0.98 & 0.18 & 0.25 & 0.04 & -49 & 5 \\
RXJ1705 & 17 05 59.6 & +36 57 34.1 & 1.23 & 218.3\tablenotemark{*}  & 41.4 & 1.65 & 0.4 & 0.17 & 0.03 & -85 & 5 
\enddata   
\tablenotetext{*}{The clusters with optically-quiescent central galaxies. These clusters have larger core radii (30-220 kpc) and significantly lower central surface brightnesses.}
\tablecomments{The parameters reported in this table are based on  two-dimensional fits of a simple 2D $\beta-$model $S_x(r) = S_0 \left(1+(\frac{r}{R_c})^2 \right)^{-\alpha}$ with eccentricity $\epsilon$ and semi-major axis position angle PA.   See text for additional details.}
\end{deluxetable*}

The amplitudes of any putative central point sources in the extended sources were statistically consistent with zero in all, as seen  in Table~\ref{table:betamodel}, with the exception of RXJ~1350+09, which has a $\sim 3\sigma-4\sigma$  X-ray point source aligned with the central radio source, a VLBI position calibrator, offset by about $1\arcsec$ from the best-fit center of the cluster. To get a stable estimator of the flux of this source required us to constrain a priori its position in a box of about $2\arcsec \times 2\arcsec$. (Fits allowing the point source position to be free failed to converge.) In this case, we obtained a $>3\sigma$ detection of a point source with a Gaussian amplitude of $77\pm^{153}_{22}$ counts pixel$^{-1}$ with a FWHM of $0.742^{+0.36}_{-0.18}$ pixels.  If we required the FWHM of this source to exceed 1.0 pixels, the corresponding Gaussian amplitude of the source was $30 \pm 8$ counts pixel$^{-1}$.  In the other seven cases, the best-fit positive point-source amplitudes differed from zero by less than $1 \sigma$. So, none of the cluster-dominated sources we found were disguised by a prominent X-ray point source. 

We note that the sources hosting quiescent central galaxies have larger X-ray core radii  ($30-220 h_{70}^{-1}$ kpc ) than those hosting emission-line central galaxies (10-30$h_{70}^{-1}$ kpc). This segregation is not that surprising, as cool-core clusters are the only clusters that are seen to host central emission-line galaxies, and their X-ray cores tend to be more compact \citepeg{2008ApJ...683L.107C,Ota2013}.  Furthermore, the clusters with the largest core radii also had the largest and most significantly non-zero ellipticities. This pattern is consistent with the notion that these clusters are unlikely to be cool-core clusters, which tend to be rounder and more relaxed.
All of the detected clusters have relatively compact core radii compared to very nearby clusters of galaxies (see the very early work by \citet{1984ApJ...276...38J}) but are rather typical of clusters studied with {\em XMM} at $z\sim0.4-0.6$ \citep{2014A&A...561A.112G}. As  mentioned in our summary, only two of these clusters in our sample had any hint of extent in the ROSAT observations, so these clusters may have been overlooked because they were somewhat more compact than typical X-ray clusters, and because the emission-line spectra of many of their central galaxies looked more like LINERS than like quiescent BCGs.

\subsection{Central AGN Variability \label{section:0905}}

The X-ray luminosity of the modest X-ray AGN in NGC~1275, the central galaxy of the nearby Perseus Cluster, has changed over the recent past \citep{2015MNRAS.451.3061F}. At least one of clusters in our sample (RXJ~0905+1840) also may have a much more prominent X-ray AGN in the not too distant past. 

Most of the {\em Chandra} fluxes for the X-ray clusters in our sample were similar (within $1 \sigma$) to expectations based on their ROSAT fluxes, and a galaxy cluster is not expected to be a variable X-ray source. Nevertheless, RXJ~0905+1840 exhibited a {\em Chandra} flux that was significantly lower than its ROSAT estimate, even taking into account the uncertainty of the ROSAT flux. 
Its flux was measured to be about one-third that expected from the ROSAT count rate, about $4\sigma$ below expectations. ($F_x(0.5-2.0~\rm{keV}) = 2.3 \times 10^{-13}$ erg s$^{-1}$ cm$^{-2}$ vs. $(6.83 \pm 1.14) \times 10^{-13}$ erg s$^{-1}$ cm$^{-2}$ in a similar aperture.) 
This cluster BCG has no discernable point source in the {\em Chandra} observation. So, it is possible that in the 1990s, during the RASS data collection phase, there may have 
been significant AGN emission contributing to its X-ray flux, but it has since faded. Its optical spectrum was classified by \citet{2006A&A...455..773V} as a Type 1 Seyfert. Interestingly, the X-ray luminosity of this group of galaxies (about $4 \times 10^{43}$ erg s$^{-1}$) is similar to the estimated weak detections and upper limits for luminosities of possible extended X-ray sources around the bright point sources in our sample (see \S5), which are between $1-10\times 10^{43}$ erg s$^{-1}$.) With regard to our {\em Chandra} observations, this cluster is like a Virgo-like cool-core cluster, with a multiphase emission-line BCG and a relatively weak Fanaroff-Riley Class 1 (FRI) radio source, but short-term fluctuations in the accretion rate onto the black hole in a cool-core cluster may cause its X-ray flux to vary.

\subsection{Spectral Analysis}

The centroids obtained from the two-dimensional morphological analyses were adopted as the centers of circular spectral extraction regions.  (See Table~\ref{table:betamodel} for the cluster-only sources.)  The background was estimated in the standard manner, by reprojecting the corresponding 
deep X-ray backgrounds supplied by the {\em Chandra}Science Center. We scaled the reprojected backgrounds by modifying the exposure time in the header to make the $9-12$ keV counts in the background spectra match the observed count rates in the same band. (For most of these observations, using the ratio of exposure times to scale the background spectra would have
been equivalent.)  To minimize contamination in the soft X-ray band, we limited the spectral fits to the energy range $0.7-7.0$ keV. We fixed the column densities to the estimated Galactic column density $N_H$ \citep{1990ARA&A..28..215D}.  With XSPEC v12.9.1  \citep{1996ASPC..101...17A}, we fit an absorption-corrected plasma model ( $phabs \times apec$ )  to obtain estimates of the normalization, X-ray temperature, and metallicity.  We fit both spectra binned to a minimum of 25 counts per energy bin and also fit unbinned spectra.   The exact results did not depend in any interesting way on the binning of the spectral data as the number of counts per spectrum was not small ($1000-3000$ counts).  We report the results of fits to the binned spectra in Table~\ref{table:Xspec}. 

We estimate temperatures
and luminosities inside an annulus between $0.3$ and $1.0$ times the scale radius $R_{2500}$, \footnote{$R_{2500}$ is the radius inside which the estimated mean density
of matter (including dark matter) is 2500 times the critical density at the same epoch as the cluster redshift.} 
we repeated the fit until the radius of extraction ($R_{2500}$) and the best-fit temperature ($T_{2500}$) were consistent with the nominal $M_{2500}--T_{2500}$ scaling relation from \citet[][,V06 hereafter]{2006ApJ...640..691V}: 
$h(z) M_{2500} = (1.25 \pm 0.15 \times 10^{14} M_\odot (T/5 {\rm keV})^{\alpha_s}$ where $\alpha_s =1.64 \pm 0.06$.
For $h=0.7$ and $kT=5$ keV and $z=0$, $R_{2500} = 501$ kpc.
Core excision avoids underestimating the X-ray temperature characteristic of the total mass of the 
cluster by excluding bright X-ray emission from a possible cool core.

The total luminosity inside $R_{2500}$ was estimated by using the spectrum as the basis for 
converting counts per second within the energy window of 0.5-7.0 keV to an X-ray luminosity integrated within
any nominal bandpass (including an effective ``bolometric'' bandpass of 0.1-12.0 keV).
However, any given spectral extraction region could be missing flux from near detector chip edges, between chip gaps or
around excised point sources.  
To estimate the total counts, we integrated the best-fit surface brightness models, taking precautions to not integrate the
model beyond the regions included in the fit. 
We corrected the net count rate from the spectra to the predicted total count rates from the best-fits
integrated within $R_{2500}$ and we found that the correction factors, or the ratios of the aperture-corrected
luminosity to that based on the spectrum alone, were not large. The ratios between fluxes from model integrations and
the fluxes from actual spectral apertures ranged from $1.04-1.47$ for the full (non-core excised) apertures
to $1.04-1.98$ for core-excised apertures.

\begin{deluxetable*}{lccccccccc} 
\tablecaption{Core-Excised Cluster Dominated X-ray Properties 0.3-1.0 R$_{2500}$  \label{table:Xspec}}
\tablehead{
\colhead{Name}           & \colhead{$T_{2500}$} & \colhead{Raw Counts}   &\colhead{Scale}       & \colhead{$R_{2500}$}    & \colhead{$Z/Z_\odot$}  & \colhead{$F_{2500}$\tablenotemark{b}} & \colhead{ACF\tablenotemark{c}} & \colhead{$L_{2500,{\rm bol}}$} & \colhead{PtSrc\tablenotemark{d}}  \\
                    & \colhead{(keV)}  & \colhead{(Net)\tablenotemark{a}}   & \colhead{(kpc/$"$)} &  \colhead{(arcsec)} &    & \colhead{($10^{-13}$ erg s$^{-1}$ cm$^{-2}$)} &               & \colhead{($10^{44}$ erg s$^{-1}$)} & \colhead{($10^{-15}$ erg s$^{-1}$ cm$^{-2}$)} } 
\startdata 
RXJ~0905  & $2.5^{+0.21}_{-0.21}$ & 568  & 2.19 & 155 & $0.42^{+0.22}_{-0.16}$ & 5.63 & 1.98 & 0.25 & $<1.5-2.5$ \\
RXJ~0927 & $3.0^{+0.3}_{-0.3}$      & 962 & 3.31 & 103 & $0.12^{+0.14}_{-0.11}$ & 8.69 & 1.18 & 1.16 & $<2.5-6$\\
RXJ~1030  & $5.2^{+0.75}_{-0.53}$ & 518 & 6.22 & 78 & $0.35^{+0.19}_{-0.16}$ & 3.34 & 1.04 & 4.24 & $<14$\\
RXJ~1119  & $3.5^{+0.7}_{-0.5}$      & 311 & 4.77 & 71 & $0.60^{+0.5}_{-0.35}$ & 2.25 & 1.70& 0.92 & $<3.5-8.1$ \\
RXJ~1131  & $3.8^{+0.40}_{-0.35}$  & 1204 &3.57 & 116 & $0.25^{+0.15}_{-0.15}$ & 11.23 & 1.22 & 1.88 & $<8.4-11$ \\
RXJ~1350 & $4.2^{+0.4}_{-0.3}$        & 6358 & 2.36  & 167 & $0.11\pm0.10$                    & 24.96 & 1.23 & 1.65& $8\pm 2$ \\ 
RXJ~1456  & $8.8^{+2.8}_{-1.5}$      & 924 & 5.50 & 99 & $<0.50$ & 7.27 & 1.45 & 5.44& $<4.8-10.2$ \\
RXJ~1705 & $6.5^{+1.1}_{-0.8}$      & 2224 & 4.22 & 114 & $0.31^{+0.19}_{-0.17}$ & 11.84 & 1.04 & 3.41 & $<5.2$ 
\enddata
\tablenotetext{a}{Net number of counts in the spectrum between $0.54$ and $7.0$ keV used to estimate X-ray temperature and metallicity, as well as the conversion between X-ray events and flux. The net source counts ranged between 60\% and 75\% of the total counts in the X-ray spectra excluding the core.}
\tablenotetext{b}{Fluxes (in $10^{-13}$ erg s$^{-1}$ cm$^{-2}$) are reported for the $0.5-7.0$ keV bandpass. Luminosities ($10^{44}~h_{70}^{-2}$ erg s$^{-1}$) are nearly bolometric, estimated by integrating to the limits of the spectral models, between 0.5-12.0 keV. All quantities in this table are computed assuming $H_0=70$ km s$^{-1}$ Mpc$^{-1}$, $\Omega_M=0.3$, and $\Omega_\Lambda=0.7$.} 
\tablenotetext{c}{ACF: Aperture correction factor}
\tablenotetext{d}{PtSrc: Point source. $3\sigma$ upper limits are based on a Gaussian, 0.5-1.0 arcseconds FWHM, X-ray fluxes between 0.5-7.0 keV, in units $10^{-15}$ erg s$^{-1}$ cm$^{-2}$. } 

\end{deluxetable*}

\section{Point Sources}

The morphologies of the six point-source-dominated cluster candidates were analyzed by fitting the masked X-ray images described in \S~3 to a model. The model consisted of a Gaussian source convolved with the {\em Chandra} ACIS point spread function and a uniform background. To determine how much residual emission could be attributed to cluster emission, we added an extended component with a fixed extent and shape and allowed the normalization to vary. Because there was usually some excess light around the point source even with the point spread function taken into account, our estimates resulted in detections with a statistical significance of $3-10\sigma$. However, our incomplete understanding of the contribution of low-level scattered light leads us to be cautious about attributing all of the excess to an extended thermal component in most (but not all)  of these cases. Therefore, the most conservative interpretations of the excesses are as upper limits. Nevertheless, it is interesting that these estimates yield X-ray luminosities consistent with the lower optical richness of the GMBCG clusters associated with these sources, compared to the ones with obvious extended emission.

\subsection{Point-Source Model}
    
    We generated custom point-source models for each source using ChaRT v2\footnote{\url{http://cxc.harvard.edu/ciao/PSFs/chart2/runchart.html}}. 
ChaRT uses the energy distribution and position of the source to simulate a set of rays through the {\em Chandra} optics. To construct the energy distribution, we extracted an X-ray spectrum for each source with circular apertures centered on the X-ray source with radii of 20 pixels ($10\arcsec$). We fit a spectrum of each source to a model including a simple absorbed power law. In each case, the best-fit spectrum model (binned in units of photons cm$^{-2}$ s$^{-1}$)  between $0.4-6.0$ keV was saved to a file and uploaded to ChaRT. Chart uses the celestial location of the point source (including the off-axis angle), and the pointing information associated with the original events file to create a collection of rays. We then used MARX \citep{2012SPIE.8443E..1AD} to project the rays into an image array, accounting for detector effects. This procedure provided the PSF  that was used in the the spatial analysis.

\subsection{Upper Limit on Extended Sources around Bright Point Sources}

\begin{figure*}
\includegraphics[width=\linewidth, angle=0]{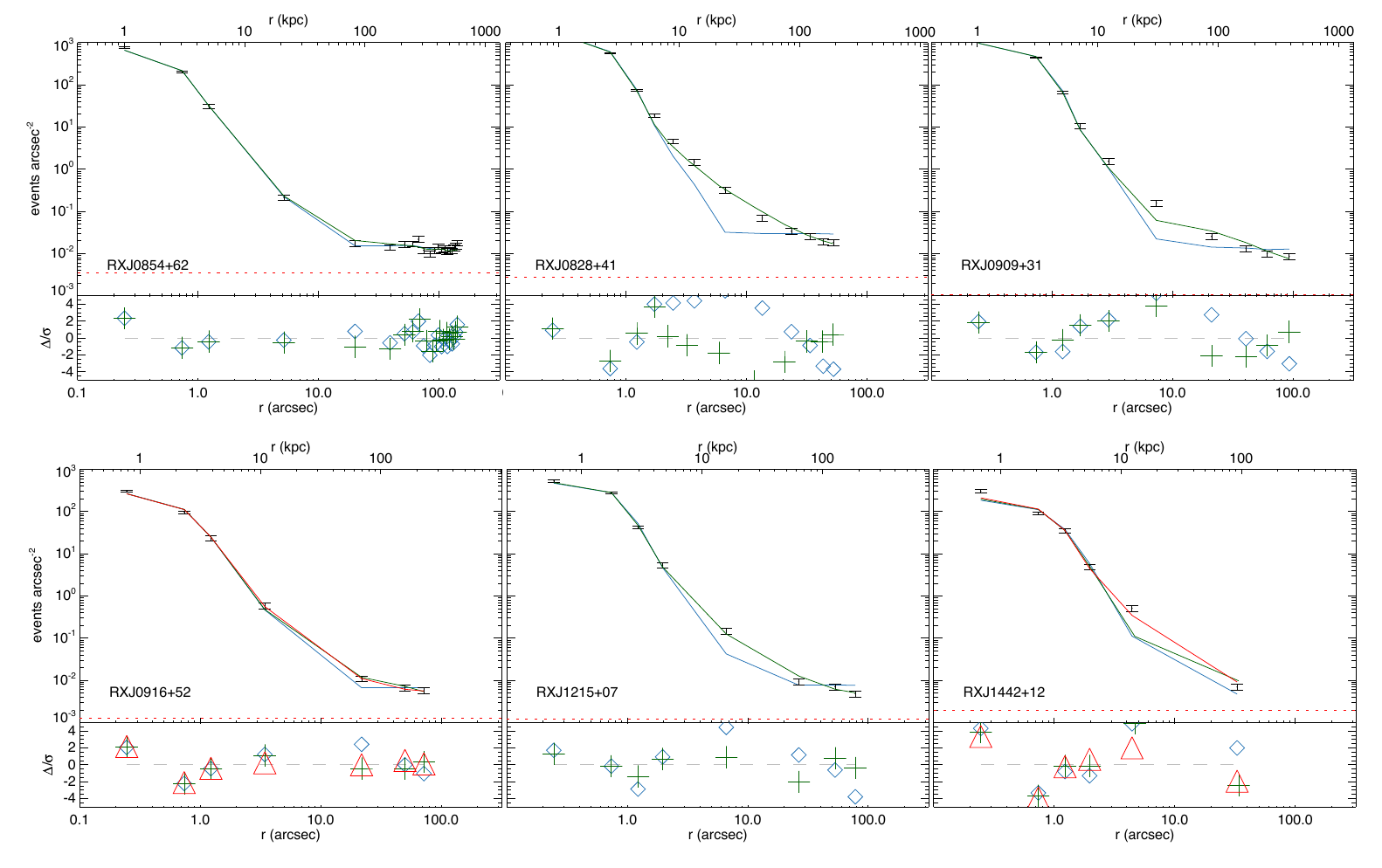}
\caption{
 Radial profiles of the six point-source-dominated sources, based on the 2D single-pixel binned X-ray images and best-fit 2D models. These graphs show the data (black points with error bars) and the models binned identically, to 50 events per radial bin. They are not the basis for the 2D fits. The red line represents the  best-fit Gaussian-blurred PSF model, binned like the data. The green line, the best-fit, shows a point source model, but with an additional extended component, a beta-model with a fixed core radius of 100 kpc, and $\beta=2/3$. The blue line is also a point-source model plus a more compact beta-model extended source, with a fixed core radius at 10 kpc, $\beta=2/3$. 
 In all cases, the model is binned identically to the data. The inferred cluster luminosity based on the best-fit extended component in each case is consistent with the estimated richness of the GMBCG cluster. The red dotted line shows the best-fit background levels inferred from the model fitting. 
\label{figure:2}
  }
\end{figure*}

We used SHERPA to obtain a best fit to the two-dimensional imaging data for a point source modeled by a very compact but finite-width Gaussian, convolved with the MARX-generated PSF. The two-dimensional data were masked to exclude regions too close to chip gaps, other point sources, and readout streaks. The Gaussian was varied to find a best-fit width, to allow for finite added blur in the observations. We used the same statistical conventions as for the 2D fitting to the clusters. To place an upper limit on a nominal extended X-ray source, we then ran a second fit with the best-fit parameters of the first fit as guesses for the Gaussian model, but in this round, we added a two-dimensional beta model with two fixed core radii  of 100 kpc and 10 kpc, in addition to a constant background. We also fixed $\alpha=1$, typical of what we found for the clusters.  To our surprise, the best-fit convolved 
PSF often did not account for all of the X-ray flux, and the best-fit amplitude of the nominal beta-model was significantly different from zero. Radial profiles of the data and the identically-binned versions of the 2D models show some evidence for this excess. (These radial profiles were created in the same manner as described in \S~4.)  These excesses should be regarded as very conservative upper limits to cluster emission, because there could be scattered light we did not account for in our PSF model. Furthermore, there could be X-ray emission from radio lobes as has been seen in radio galaxies \citep{2003ApJ...584..643D,2005ApJ...626..733C}. 
In the radial profiles of the data and the models in Figure~\ref{figure:2}, we see that the case for an extended component is sometimes strong, while in others the extended signal could arise from large-angle scattering of X-rays unaccounted for in the MARX models.  

Given the very low number of excess counts spread over a relatively large area, we cannot reliably obtain an X-ray spectrum for the extended component, so estimating $R_{2500}$ for these candidate clusters was not possible. But to give an approximate scale, we provide the estimate of what $R_{2500}$ in kiloparsecs would be for an X-ray temperature of 2 keV in each case in Table 4, along with the flux and luminosity estimates inside that radius for the clusters around these AGN if 100\% of the extended flux is indeed from hot gas.  To get an idea of whether the luminosities were similar to those of clusters, groups, or even individual galaxies, we estimated the implied X-ray fluxes of the extended component within a metric radius of 300 kpc. We used the best-fit amplitudes for a fixed-shape beta-model ($r_c=100$ kpc was assumed for all but two sources for which the $r_c=10$ kpc fit was marginally better) and integrated that best-fit model to estimate the total flux and luminosity inside 300 kpc and inside the nominal $R_{2500}$ for a cluster of 2 keV.  

The most convincing case of extended emission around an AGN was RXJ~0828+41, and the estimated luminosity of this component was $\sim 3.6\times10^{43}$ erg s$^{-1}$, not so  different from that of the Virgo Cluster. The amplitude of a core radius $r_c=10$ kpc beta model had an estimated statistical significance of $\sim 10\sigma$ for this source (Table\ref{table:qsoextlimits}). The most dubious detection was for RXJ~0854+62, with an estimated component luminosity of $\lesssim 10^{42}$ erg s$^{-1}$, representing only a 3-$\sigma$ statistical detection. The other AGN-dominated sources showed plausible excesses at radii of 10-100 kpc, corresponding to  $L_x \sim 1-10 \times 10^{43}$ erg s$^{-1}$. The levels of these excesses are consistent with those of massive group or poor cluster luminosities near the optical richness of the GMBCG systems (see \S\ref{section:comparisons}.)

\begin{deluxetable*}{lcrrccrrcc}
\tablecaption{Estimated Extended Source Counts around X-Ray AGN in BCGs \label{table:qsoextlimits}}
\tablehead{
\colhead{Name} & \colhead{Scale} & \colhead{Core $R_c$} & \colhead{$R_{2500}$}\tablenotemark{a}  & \colhead{Amplitude}   & \colhead{Background} & \colhead{Counts}   & \colhead{Counts}  & \colhead{$F_x$ (0.5-7.0) ($<R_{2500}$)} & \colhead{$L_x$}  \\ 
\colhead{} & \colhead{(kpc/$"$)} & \colhead{(kpc)} &  \colhead{(kpc)} & \colhead{(counts per pixel)}  & \colhead{(counts per pixel)}   & \colhead{($<R_{2500}$)}  & \colhead{($<300$kpc)} & \colhead{($10^{-13}$  erg s$^{-1}$ cm$^{-2}$)} & \colhead{($10^{43}$ erg s$^{-1}$) }  }
\startdata
RXJ~0828 & 3.63 & 10          & 271                 & 0.507     $\pm$ 0.045    & 0.0028     $\pm$ 0.0004         & 325           $\pm$ 29          & 334                   & 2.37                                   & 3.6            \\
RXJ~0854 & 4.10 & 100         & 265                 & 0.0031         $\pm$ 0.0011    &    0.000354        $\pm$ 0.00016     & 50   $\pm$ 16          & 54                    & 0.33                                   & 0.7            \\
RXJ~0909 & 4.16 & 100         & 264                 & 0.013          $\pm$ 0.0014   & 0.00105    $\pm$ 0.0002         & 203      $\pm$ 22          & 226                   & 5.05                                   & 12             \\
RXJ~0916 & 3.17 & 100         & 277                 & 0.0036           $\pm$ 0.0008   & 0.0008     $\pm$ 0.0002         & 100        $\pm$ 22          & 106                   & 2.95                                   & 3          \\
RXJ~1215 & 2.41 & 10          & 284                 & 0.08              $\pm$ 0.012    & 0.001      $\pm$ 0.0014         & 119      $\pm$ 18          & 125                   & 2.37                                   & 1.2            \\
RXJ~1442 & 2.80 & 100         & 280                 & 0.0073            $\pm$ 0.001    & -0.00066   $\pm$ 0.00018        & 285       $\pm$ 39          & 297                   & 16.4                                   & 12             
\enddata
\tablenotetext{a}{$R_{2500}$ estimated from relation in V06, assuming $kT \sim 2$ keV and cosmological parameters $H_0=70$ km s$^{-1}$ Mpc$^{-1}$, $\Omega_M=0.3$, $\Omega_\Lambda=0.7$.}
\end{deluxetable*}

\section{Discussion of Classification Results}

High-resolution {\em Chandra} data allowed us to trivially distinguish the X-ray sources in our sample dominated by extended emission from those dominated by point-source emission. The X-ray source modeling presented in this paper was conducted to characterize the emission in the clearly extended sources and to provide a basis for deriving upper limits for point sources and extended emission in the two categories. 
Here we discuss our results, aggregated by the classification information available prior to the {\em Chandra} observations.  We also include some notes on individual sources. 

\subsubsection{Extended ROSAT Sources}

Three of the fourteen sources in our sample (RXJ~0905+1840, RXJ~0909+3105, and RXJ~1350+0940) were identified as having a probability greater than 99.7\% of being extended in the RASS Bright Source Catalog (BSC) \citep{Voges_1999} but were not subsequently identified as clusters. One of the sources, RXJ~0909+31, was classified spectroscopically as an AGN by us and by \citet{2003A&A...406..535Z}, and we confirmed in this work it is almost completely dominated by the X-ray AGN with $<1\%$ of the signal coming from a possible extended component. The other two were identified as probable clusters in the RASS optical identification  by  \citet{2003A&A...406..535Z}, but were not listed in  MCXC. It is now clear that others certainly knew at least RXJ~1350+09 was a cluster based on its public {\em Chandra} abstract. We confirm both RXJ~0905+1840 and (unsurprisingly) RXJ~1350+0940 as clusters. 

The other 11 targets had no significant detectable extent seen in the RASS, just as the RASS BSC reports for the Phoenix Cluster itself.  With {\em Chandra} we found at least 6 of the 11 X-ray ROSAT point sources to be unambiguous X-ray clusters, two-thirds of them cool-core clusters, for a total of 8 well-detected clusters out of the sample of 14. The remaining six have X-ray emission dominated in each case ($>90\%$) by an X-ray AGN point source. These AGN plausibly reside in X-ray groups with faint X-ray luminosities commensurate with the optical richnesses estimated for their GMBCG cluster hosts by \citet{Hao_2010}.

\subsection{Classic Cool Cores (CC)}  

Six of the fourteen X-ray sources represented potential cool-core clusters, based on the spectra of their BCGs. All six of these sources turn out to be extended {\em Chandra} sources, suggesting that the spectroscopic identification of the characteristic emission-line spectrum associated with cool-core clusters \citep{Heckman_1989} is an efficient way to distinguish X-ray clusters from X-ray AGNs when X-ray spatial resolution is insufficient to distinguish the two. To our surprise, we failed to detect any evidence for a central X-ray AGN in any of these sources except for RXJ~1350+0940. They are all radio sources with a fairly wide range of radio luminosities, ranging from relatively radio-quiet (RXJ~0905+1840 at $L_{1.4Ghz} =4.3\times 10^{23}$ W Hz$^{-1}$) to very radio-loud (RXJ~1030+5132 at $L_{1.4Ghz} =9.0\times 10^{25}$ W Hz$^{-1}$). 

Brief notes about the cool-core clusters: 
\begin{enumerate}

\item RXJ~0905+1840 is a low-luminosity radio source in a cluster of low optical richness hosting a BCG with very bright, LINER-class emission lines.  

\item RXJ~1030+5132, a radio-loud, high-$L_x$ source, was named as an X-ray source in a table embedded in \citet{2003MNRAS.339..913E}. Edge et al. cites \citet{2000A&AS..144..247C} as identifying RXJ~1030+51 as a cluster. An inspection of that paper shows Caccianiga et al. identified RXJ1030 as a possible broad-line AGN, so it appears that Edge et al. may have been the first to identify it as an X-ray cluster. Nevertheless, this cluster is not in the MCXC nor is it included in flux-limited surveys. It hosts a $z=0.5185$ flat-spectrum quasar inhabiting a BCG with very bright emission lines. It is the most X-ray luminous of our CC cluster candidates with a bolometric X-ray luminosity of almost $10^{45}$ erg s$^{-1}$. This cluster is the most Phoenix-like cluster in our sample, but has considerably lower X-ray luminosity, and, as can be seen in Table~3, lower X-ray temperature as well. 

\item RXJ~1119+0900 is a $z=0.331$ cluster discovered in the optical by \citet{2003AJ....125.2064G} and it hosts a BCG with bright LINER-class emission lines. 

\item RXJ~0927+5327 is ZwCL2379, a Zwicky cluster hosting a BCG with emission lines. 

\item RXJ~1131+3344 is a radio galaxy with bright emission lines.  

\item RXJ~1350+0940 ($z=0.1325$), is our archival source  (Cycle 12 Proposal ID 13800738, PI Alastair Edge).  It is a cool-core cluster. We note that its quoted NED redshift of 0.09 (photometric, from \citet{2003AJ....125.2064G}) is very different from its SDSS spectroscopic redshift of 0.1325. In this work we assume $z = 0.1325$.  It was the only composite source in our sample in which the extended X-ray emission was the dominant source of X-rays. We detect a faint X-ray point source aligned with the central radio source, which is a VLBI calibrator source.

\end{enumerate}

\subsection{AGN-Dominated Clusters (AGN)}.  

Four targets (RXJ~0854+6218, RXJ~0909+3105, RXJ~0916+5238, and RXJ~1442+1200) represent cluster candidates that were identified in advance of our {\em Chandra} observations as likely to be dominated by X-ray emission from  an AGN residing in the BCG.  Their estimated X-ray luminosities  were large ($>\rm{few} \times 10^{44}$ erg~s$^{-1}$) so even a subdominant cluster or group contribution could still be interestingly high.  RXJ~1442+1200 has an intriguing, off-center blue source in the SDSS image that could be the source of some of the X-ray emission. We confirmed all of these sources to be almost completely ($>90$\%) dominated by X-ray AGN point-like sources. 

We detected or placed limits on faint extended X-ray emission around each of these sources. (See Table~\ref{table:qsoextlimits} for a summary of the search for extended emission.) Some of this extended emission could be stray light from the bright point source, scattered by the X-ray optics, and some if it could be from thermal emission from gas around the source, and some could even be emission related to real X-ray structure associated with X-ray photons emitted or scattered by radio lobes \citep[e.g., ][]{2003ApJ...584..643D}. The faint level (55-300 net counts) of the off-axis signal makes it impossible to definitively discern the difference between thermal and nonthermal spectra, especially as low-level scattering from the AGN certainly accounts for some of the events in the detector regions surrounding the AGN. But, even after carefully modeling the spatial distribution of a point-source component, we were also unable to categorically rule out group-scale extended emission below a formal statistical $3\sigma$ detection in any of these sources.

\subsection{Clusters with Optically Quiescent BCGs (Q)}  
The remaining 4 of the 14 cluster candidates (RXJ~0828+4153, RXJ~1215+0732, RXJ~1456+3021,  and RXJ~1705+3657) have BCGs with optical spectra showing little to no emission-line components and which are characteristic of old stellar populations. RXJ~0828 and RXJ~1215 have relatively strong radio sources, while the other two have only upper limits from the FIRST survey. 

Our {\em Chandra} observations revealed that half of these candidates were clusters and half were AGNs. The most optically rich cluster in this group (RXJ~1705+36) is also an X-ray cluster; the most optically poor cluster (RXJ~1215+07) appears to be an X-ray AGN with faint X-ray extended emission commensurate with its low optical richness. The other two optical clusters have moderate optical richnesses, one with and one without an X-ray AGN. Interestingly, the only two sources in our entire sample with very weak to absent radio sources (RXJ~1456+30 and RXJ~1705+36) turned out to be  X-ray clusters (Table \label{table:betamodel}). The two quiescent BCGs with luminous radio sources were revealed to be bright X-ray AGN point sources with only very faint extended emission (Table~\ref{table:qsoextlimits}). So at least among quiescent BCGs, strong radio emission may distinguish between X-ray sources that are X-ray luminous clusters (radio-quiet) and X-ray point sources in groups (radio-loud). Optical richness alone is not a particularly reliable indicator because of the large intrinsic scatter with respect to X-ray luminosity, but one would expect that the most optically-rich clusters to have strong X-ray emission from their hot gas as well.

These quiescent BCGs may be representative of unidentified X-ray sources associated with quiescent galaxies associated with radio sources. We also hoped that follow-up of these sources might have provided examples of cool-core clusters that lack emission-line BCGs. One famous example of a cool-core cluster with stringent limits on H$\alpha$ emission is Abell 2029 \citep{Jaffe2005}, an unusual X-ray cluster seemingly perched on the border between strong cool-core clusters and those with higher central gas entropy more easily stabilized by conductive processes \citep{2011ApJ...740...28V, VoitDonahue2015}. It has been speculated that Abell 2029 has unusually low gas turbulence compared to the escape speed from its gravitational potential \citep{2018ApJ...868..102V}. So our prior expectation (or even hope) was that most of the X-ray emission in these quiescent BCGS may be actually from hot gas because the SDSS spectroscopy for these BCGs lack AGN signatures such as strong emission lines or strong blue continua. But only two of them emerged as clear X-ray clusters, and those two were the sources with weak to absent radio sources. 

One then might be tempted to generalize from this outcome that  X-ray sources around optically quiescent galaxies without radio sources are likely X-ray clusters or groups, while sources associated with quiescent galaxies with powerful radio sources are X-ray AGNs, but this conclusion would be incorrect, and could lead to surveys missing very interesting and important objects. An Abell 2029-type cluster would present a relatively strong radio source inside a compact, peaked X-ray surface brightness profile.  So, an X-ray cluster selection algorithm that categorically rejects optically quiescent BCGs with powerful radio sources could miss the twins of clusters like Abell 2029. Based on our small sample, many of optically quiescent, compact X-ray sources would likely turn out to be AGN-dominated with perhaps some faint extended emission. There is, sadly, no counterpart in our sample to the extreme Abell 2029 BCG. Both of the quiescent galaxies with luminous radio emission in our sample are also luminous X-ray AGN, which Abell 2029's BCG is not. This outcome is consistent with the idea that twins of Abell 2029 are relatively rare because it may be in a relatively short-lived state of quiescence.  Abell 2029 could also be rare if the BCG in Abell 2029 has an unusual steep gravitational potential. Studies of the velocity dispersion of stars in the BCG of Abell 2029 show that not only is the velocity dispersion high, but it increases with radius \citep{1979ApJ...231..659D} which is indeed relatively unusual in BCGs \citep{2018MNRAS.477..335L}.

Abell 2029 and systems like it are certainly worth seeking and receiving further regard, so future X-ray surveys should take care not to exclude powerful radio sources in optically quiescent galaxies from their cluster candidates without careful follow-up.

\section{Other Archival {\em Chandra} Data}

Three of the sources on the complete list of 25 potential X-ray cluster candidates that we compiled were not observed by us with {\em Chandra} because there were already existing observations. We included one of these already in this analysis since the data were acquired in 2012, but had not yet been published (RXJ1350+09).   Another candidate was the radio galaxy 3C~219, which has a bright central point source and extended emission. The extended emission around this source is at least partly related to inverse Compton scattering by radio plasma in the radio lobes \citep{2003MNRAS.340L..52C,2005ApJ...626..733C}. The radio source 3C~219 is in an optically identified cluster of galaxies, and distinguishing extended X-ray emission from intracluster gas vs. the X-ray lobes is not trivial \citep{2005ApJ...626..733C}. The third source, ES~0927+50.0 is a $z=0.186742$ QSO with very bright optical emission lines. It has a Chandra/HRC archival observation that shows little evidence of extended emission and it would be classified by us as an X-ray AGN.

\section{Comparisons with Typical Clusters \label{section:comparisons}}

\begin{figure}
\includegraphics[width=\linewidth, angle=0]{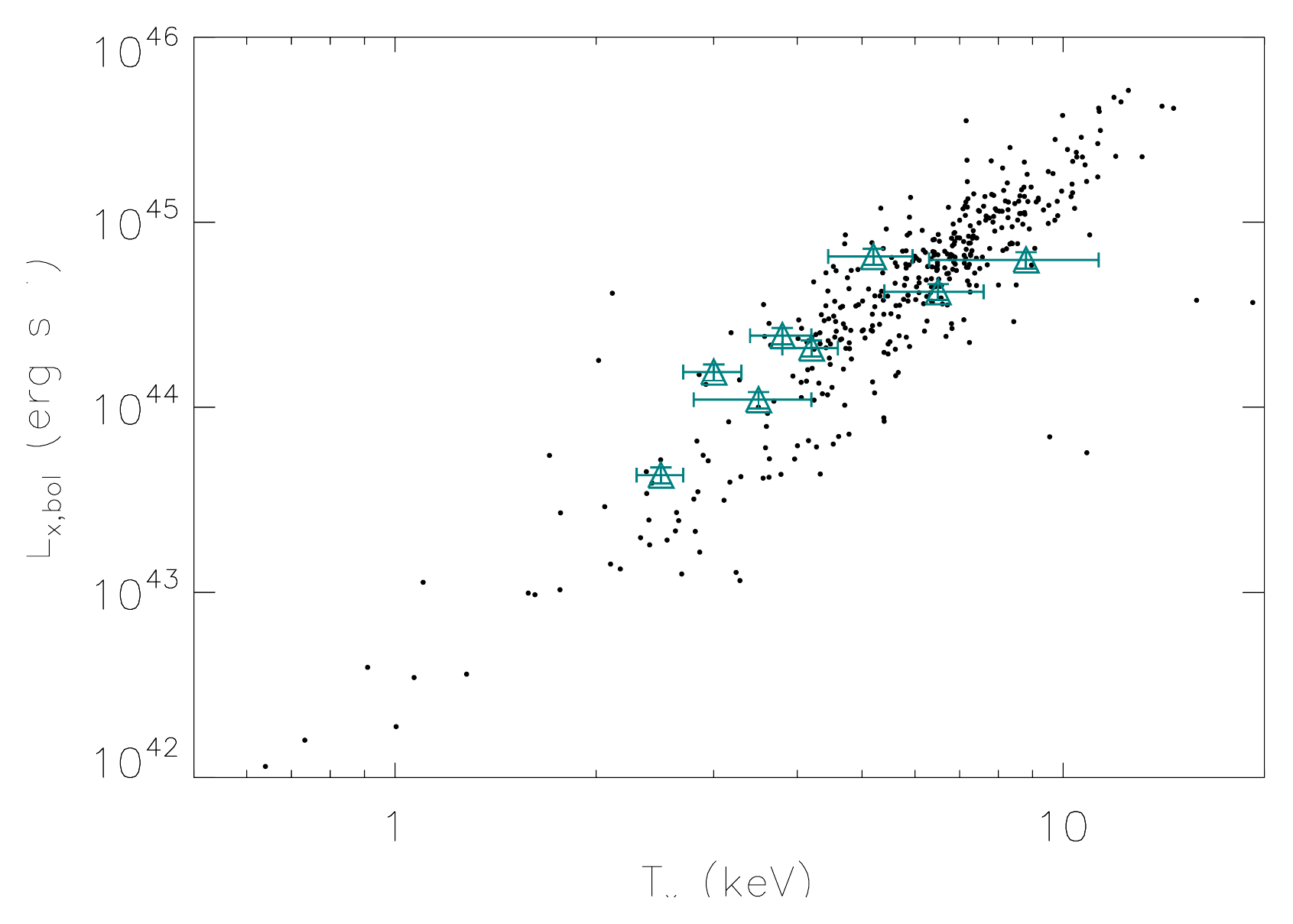}
\caption{
 The data for the X-ray clusters characterized in this paper are plotted with teal triangles with error bars. We plot the core-excised (0.3-1.0 $R_{2500}$) bolometric luminosities and corresponding X-ray temperatures for the extended sources in our sample. A comparison sample of a {\em Chandra} archival sample of  clusters (ACCEPT2) with $R_{2500} < 5\arcmin$, 382 clusters with single aperture spectra (black points, no error bars) and a sample of 22 clusters with temperatures based on emission-weighted projected temperatures are plotted with x symbols.  The X-ray temperatures and core-excised luminosities of the clusters in our sample are  typical of clusters in the {\em Chandra} archive.  
\label{figure:LxTxtest}
 }
\end{figure}

Figure~\ref{figure:LxTxtest} newly measured X-ray properties of the clusters in our sample show that they tend to have typical luminosities for their core-excised gas temperature, when compared 
to ACCEPT2 clusters of similar temperature (M. Donahue et al, j2020, in preparation). ACCEPT2 is an archive-limited sample of clusters of galaxies for which the global X-ray properties of luminosity and temperature are uniformly measured with techniques identical to those we used in this work.  This result is robust to whether we used total or core-excised properties. The two most luminous clusters in our sample are somewhat fainter than the mean in ACCEPT2 for their estimated temperature. These two clusters as noted previously are also the two clusters (RXJ~1456 and RXJ~1705) with the lowest BCG radio luminosities, two of the four optically quiescent BCGs.  Their core radii are the largest in the sample, but perhaps a little more compact than those of nearby clusters of galaxies. 

\begin{figure}
    \includegraphics[width=0.4\textwidth]{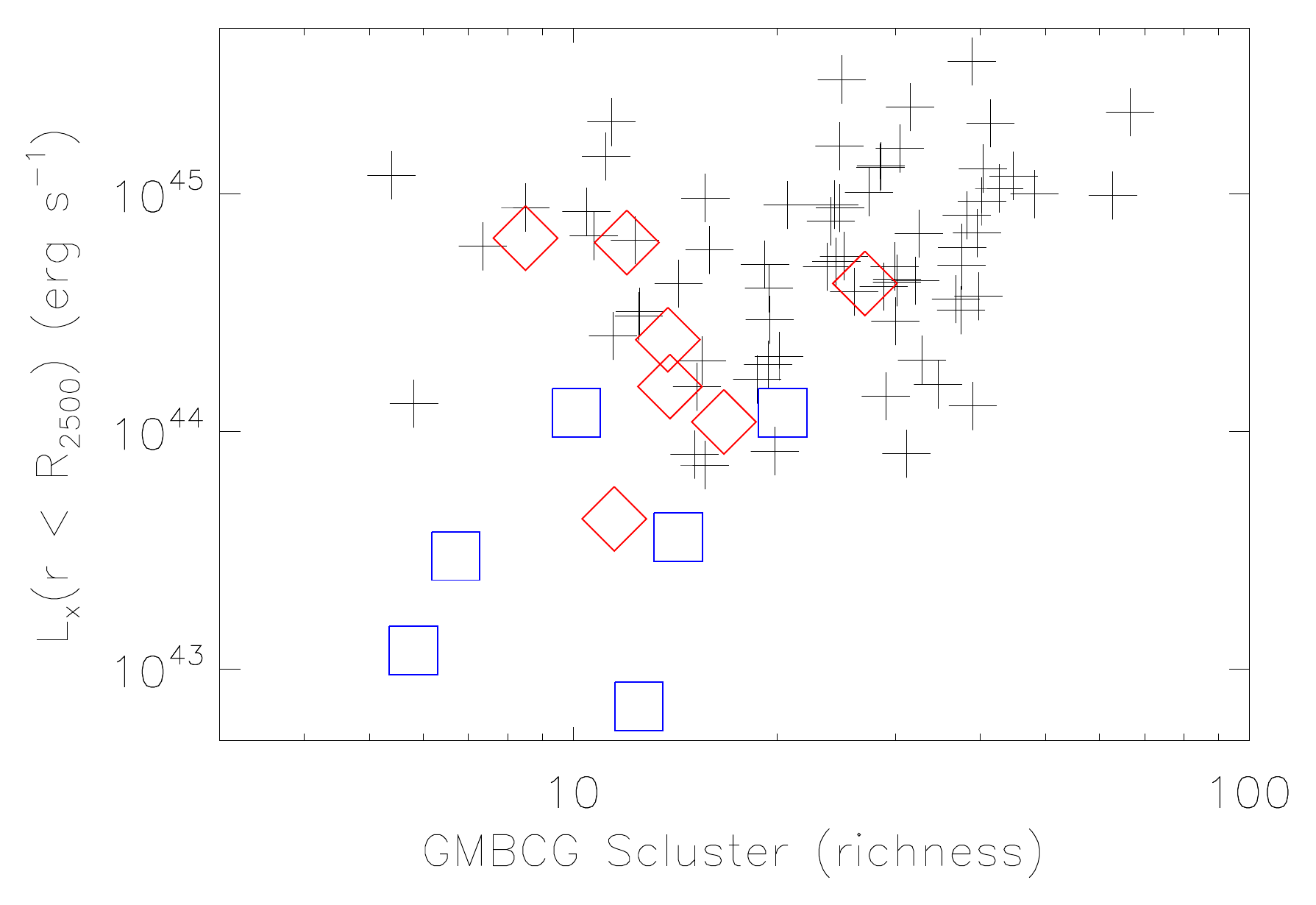}
    \caption{X-ray luminosities  (rest-frame 0.5-7.0 keV, total estimated luminosity inside $R_{2500}$)  and GMBCG optical richness (``S\_cluster'') from \citet{Hao_2010} of 69 ACCEPT2 clusters with centroids within $15\arcsec$ of the BCG locations in the GMBCG catalog and $z>0.04$. The extended X-ray sources in our sample are plotted with red diamonds. The estimate of $R_{2500}$  around the X-ray AGN depends on the source redshift, and ranges from 265-280 kpc. Estimates based on apertures with radii of 300 kpc would not be visibly different.The GMBCG optical richnesses and X-ray luminosities within based on estimates of extended X-ray emission around the X-ray quasars in our sample are plotted in blue. In this comparison, the optical and X-ray luminosities of the extended sources are typical of clusters previously observed with {\em Chandra}. The X-ray estimates of the maximum luminosities of the extended components around the X-ray AGN in our sample are fainter than the luminosities of the X-ray clusters in our sample, but are consistent  with the richness and X-ray luminosities of groups and poor clusters with $T<2$ keV, with increased scatter. 
    \label{figure:LxS}}
\end{figure}

In Figure~\ref{figure:LxS} we plot cluster richness ($S_{\rm cluster}$ from the SDSS cluster catalog) versus total cluster X-ray luminosities inside an estimated scaled radius $R_{2500}$ for our candidates and also the same measurements for 69 ACCEPT2 clusters of galaxies that are in both the {\em Chandra} archive and the GMBCG catalog and have redshifts $>0.04$, so the luminosities are based on direct measurements inside $R_{2500}$, without extrapolation.  The optical--X-ray relation shows statistical correlation, but very large scatter. 
The X-ray clusters discovered during this search to have dominant extended emission have X-ray luminosity--optical richness ratios typical of the ACCEPT2 sample. In contrast, plotting the estimated X-ray group/poor cluster luminosities around the point sources versus SDSS cluster richness shows that they sample lower X-ray luminosity limits of the observed relation. The X-ray luminosities inferred for the extended excess emission around the AGN are typical of groups, and estimates we report for these GMBCG systems are consistent with X-ray luminosities typical of clusters and groups of comparable optical richness.  

We suspect that these clusters were missed in RASS follow-up classifications not because they were under-luminous but because they may have been a little more compact and (in some cases) somewhat more distant.
Nearly all of these clusters exhibit relatively compact cores ($\sim 10-30$ kpc) in their X-ray surface brightness maps.  Such small cores are generally associated with clusters having multiphase gas and/or a strong radio source. 

\section{Summary}

Obtaining a complete sample of X-ray-selected clusters is not easy.  The ROSAT survey of the 1990s remains the most recent and open-access all-sky X-ray survey.  
Because of its large field of view (up to a square degree, with the central 15' by 15' with adequate image quality), its collection
of pointed observations and its all-sky scan survey (RASS) has been a rich treasure trove of new X-ray sources for a generation of X-ray astronomers.  ROSAT's angular resolution was not quite sufficient to reliably distinguish X-ray clusters from more point-like X-ray AGNs at moderate redshifts, especially compact X-ray clusters. Because of that limitation, one of the
most luminous and extraordinary clusters known today, the Phoenix Cluster, was missed by X-ray observers, and may have been identified as a simple quasar in earlier classifications.

In an attempt to uncover any other hidden diamonds in the ROSAT Bright Source Catalog, we cross-correlated that catalog with a large sample of known BCGs from the Sloan Digital Sky Survey. This cross-correlation yielded at least 25 sources previously unidentified as clusters, and we followed up with short {\em Chandra} observations of 13 of these sources and an analysis of one archival observation. This strategy proved to be an excellent method of
revealing X-ray clusters that may have been too compact or faint for ROSAT to unambiguously identify as extended sources.  
We detected six previously unknown X-ray luminous clusters, four of which are 
strong cool-core clusters with active emission-line BCGs and two of which are very hot and luminous with optically-quiescent BCGs. The remaining eight sources were dominated
by point-source emission from AGN. Analysis of the {\em Chandra} observations, including modeling the point sources, showed that the  galaxy groups in which these systems reside are have estimated X-ray luminosities and upper limits to their X-ray luminosities consistent with their optical richnesses, which are also similar to groups and poor clusters. 
We compared those systems with an archival sample of clusters 
of galaxies with {\em Chandra} observations (ACCEPT2) and found that their X-ray and optical properties are typical of other clusters observed with the {\em Chandra} X-ray Telescope. 
The core radii of these clusters are somewhat more compact than typical clusters, which may have contributed to their misclassification in the original
ROSAT cluster searches.  

 By observing a sample of optically-selected BCGs cross-identified with X-ray sources (regardless of extent), we obtained an interesting sample of
 previously unstudied and, in many cases, unknown X-ray clusters. We also found hints of possible X-ray group-gas emission around X-ray AGN, with X-ray luminosities consistent with
 their estimated optical richnesses. Finally, we presented evidence for a luminous X-ray AGN which has turned off inside a Virgo-like cluster within the last
 20 years or so.

\acknowledgments
This work was supported by two Smithsonian Astrophysical Observatory/National Aeronautics and Space Agency Chandra Science Center Guest Observing grants (SAO/NASA GO4-15124X, GO5-16132X). This research has also made use of data obtained from the Chandra Data Archive and the Chandra Source Catalog. This research made extensive use software provided by the Chandra X-ray Center (CXC) in the application packages CIAO, ChIPS, and SHERPA. These data were also analyzed in support of Kelsey Funkhouser's PhD candidacy second-year project at Michigan State University.
This research has made use of the NASA/IPAC Extragalactic Database (NED) which is operated by the Jet Propulsion Laboratory, California Institute of Technology, under contract with the National Aeronautics and Space Administration.



\bibliographystyle{yahapj}
\bibliography{PhoenixSearch}
\end{document}